\documentclass[a4paper,nolineno,USenglish,cleveref,autoref,thm-restate]{socg-lipics-v2021}
 \category{Media Exposition} 
 
\usepackage{amsthm}
\usepackage{amsmath}
\usepackage{amssymb}
\usepackage{amsfonts}
\usepackage{epsfig}

\usepackage{url}

\usepackage{enumitem}
\usepackage{algorithm}
\usepackage{algorithmic}

\usepackage{latexsym}
\usepackage{psfrag}
\usepackage{enumerate}
\usepackage{here}
\usepackage{esvect}
\usepackage[update,prepend]{epstopdf}
\usepackage{anyfontsize}
\usepackage{epsfig}
\usepackage{epsf}
\usepackage{fancyvrb}

\usepackage{hyperref}

\usepackage{rotating}

\usepackage{microtype}%if unwanted, comment out or use option "draft"

\makeatletter
\DeclareFontFamily{U}{tipa}{}
\DeclareFontShape{U}{tipa}{m}{n}{<->tipa10}{}
\newcommand{\arc@char}{{\usefont{U}{tipa}{m}{n}\symbol{62}}}%

\newcommand{\arc}[1]{\mathpalette\arc@arc{#1}}

\newcommand{\arc@arc}[2]{%
  \sbox0{$\m@th#1#2$}%
  \vbox{
    \hbox{\resizebox{\wd0}{\height}{\arc@char}}
    \nointerlineskip
    \box0
  }%
}
\makeatother

\newcommand{\later}[1]{{}}
\newcommand{\old}[1]{{}}
\long\def\ignore#1{}

\title{Visualizing WSPDs and their applications\thanks{Research on this paper is supported by the NSF award CCF-1947887.}}

\titlerunning{Visualizing WSPDs and their applications}

\author{Anirban Ghosh}
{School of Computing,
University of North Florida, USA}
{anirban.ghosh@unf.edu}
{0000-0003-0130-5968}
{}
\author{FNU Shariful}
{School of Computing,
University of North Florida, USA}
{n01501509@unf.edu}
{0000-0002-9561-3613}
{}
\author{David Wisnosky}
{School of Computing,
University of North Florida, USA}
{n01153911@unf.edu}
{0000-0001-5463-1949}
{}

\authorrunning{A.~Ghosh, F.~Shariful, and D.~Wisnosky}
\Copyright{A.~Ghosh, F.~Shariful, and D.~Wisnosky}


\EventNoEds{2}
\EventLongTitle{38th International Symposium on Computational Geometry (SoCG 2022)}
\EventShortTitle{SoCG 2022}
\EventAcronym{SoCG}
\EventYear{2022}
\EventDate{June 7--10, 2022}
\EventLocation{Berlin, Germany}
\ArticleNo{XX}  % <-- This will be filled in by the typesetters

\begin{document}

\maketitle

\begin{abstract}
Introduced by Callahan and Kosaraju back in 1995, the concept of well-separated pair decomposition (\textsf{WSPD}) has occupied a special significance in computational geometry when it comes to solving distance problems in $d$-space. We present an in-browser tool that can be used to visualize \textsf{WSPD}s and several of their applications in $2$-space. Apart from research, it can also be used by instructors for introducing \textsf{WSPD}s in a classroom setting.  The tool will be permanently maintained by the third author at \url{https://wisno33.github.io/VisualizingWSPDsAndTheirApplications/}. %Since  the entire codebase is publicly available  on \texttt{GitHub}, our tool can be easily tweaked and/or extended for related applications.
\end{abstract}

\section{Introduction}  

Let $P$ and $Q$ be two finite pointsets in $d$-space and $s$ be a positive real number. We say that $P$ and $Q$ are well-separated with respect to $s$, if there exist two congruent disjoint balls $B_P$ and $B_Q$, such that $B_P$ contains the  bounding-box of $P$, $B_Q$ contains the  bounding-box of $Q$, and the distance between $B_P$ and $B_Q$ is at least $s$ times the common radius of $B_P$ and $B_Q$. The quantity $s$ is referred to as the \emph{separation ratio} of the decomposition. Using this idea of well-separability, one can define a well-separated decomposition of a pointset (\textsf{WSPD})~\cite{callahan1995decomposition} in the following way. Let $P$ be a set of $n$ points in $d$-space and $s$ be a positive real number. A well-separated pair decomposition for $P$ with respect to $s$ is a collection of pairs of non-empty subsets of $P$, $\{A_1,B_1\},\{A_2,B_2\},\ldots,\{A_m,B_m\}$ for some integer $m$ (referred to as the size of the \textsf{WSPD}) such that 
\begin{itemize}
    \item for each $i$ with $1 \leq i \leq m$, $A_i$ and $B_i$ are well-separated with respect to $s$, and
    \item for any two distinct points $p,q \in P$, there is exactly one index $i$ with $1 \leq  i \leq m$, such that  $p \in A_i, q \in B_i$, or $p \in B_i, q \in A_i$. 
\end{itemize}
Note that in some cases, $m=C(n,2) = \Theta(n^2)$. Refer to \cite{har2011geometric,narasimhan2007geometric,smid2018well} for a detailed discussion on \textsf{WSPD}s and their uses. In this work, we consider \textsf{WSPD}s in $2$-space only. Our implementations are based on the  algorithms presented in the book by Narasimhan and Smid~\cite[Chapters 9 and 10]{narasimhan2007geometric}. These algorithms were originally presented in~\cite{callahan1993faster,callahan1995algorithms,callahan1995decomposition} by  Callahan and Kosaraju.

\section{Algorithms implemented}
We have implemented the algorithms using the \textsf{JSXGraph} library. Some code segments have been borrowed from the tool presented in \cite{anderson2021interactive}.

%In this section we briefly describe the algorithms we have implemented in our applet. 

\subsection{Constructing WSPDs}

   Given a pointset $P$ and a positive real number $s$, a \textsf{WSPD} of $P$ can be constructed using a split-tree. Our implementation is based on the naive quadratic time approach presented in~\cite{narasimhan2007geometric}.  It accepts $P$ and  $s$, and returns the \textsf{WSPD} pairs in the \textsf{WSPD} decomposition. Refer to Algorithm~\ref{alg:wspd}. An  advanced linearithmic construction is also presented in~\cite{narasimhan2007geometric}. 
   
   \textbf{Notations.} Let $x$ be a split-tree node. Let $S_x$ denotes the points stored in the subtree rooted at $x$ and $R(x)$
denotes the bounding-box of $S_x$. Further, $L_{max}(R(x))$ denotes the length of the longer side of $R(x)$.
    
    \begin{algorithm}[H]\caption{: \textsc{ConstructWspd}($P,s>0$)}
    
  %   \internallinenumbers 

 \begin{enumerate}\itemsep4pt

        \item  Construct a split-tree $T$ on $P$ in the following way: 
        
\medskip
        
         If $|P|=1$, then the split-tree consists of one single node that stores that point. 
        Otherwise, split the bounding-box of $P$ into two rectangles by cutting the longer side of the bounding-box into two equal parts. Let $P_1$ and $P_2$ be the two subsets of $P$ that are contained in these two new rectangles. The split-tree for $P$ consists of a root having two subtrees, which are recursively defined for $P_1$ and $P_2$.
        
        \item For each internal node $u$ of $T$, %call \textsc{FindPairs$(v,w,s)$},
        find \textsf{WSPD} pairs using $v$ and $w$,  the left and right child of $u$, respectively, in the following way:
\begin{enumerate}
\item  Compute $S_v$, $S_w$, $L_{max}(R(v))$ and $L_{max}(R(w))$. 

\item If $S_v,S_w$ are well-separated with respect to $s$, then node pair $\{v,w\}$ is a \textsf{WSPD} pair.

Otherwise, if $L_{max}(R(v)) \leq L_{max}(R(w))$, recursively find \textsf{WSPD} pairs using $v$, $\textsc{LeftChild}(w)$ and then recursively find \textsf{WSPD} pairs using $v$, $\textsc{RightChild}(w)$.

Else, recursively find \textsf{WSPD} pairs using $\textsc{LeftChild}(v), w$,  and then recursively find \textsf{WSPD} pairs using $\textsc{RightChild}(v), w$.
    \end{enumerate}
    
    \end{enumerate}

 	\label{alg:wspd}
 	\end{algorithm}

% \begin{algorithm}[H]\caption{: \textsc{FindPairs}($v,w,s$)}
% \textbf{Notations.} 
% \medskip

% \label{alg:findpairs}
% \end{algorithm}

    %However, using an involved construction, this task can be completed in $O(n\log n)$ time~\cite{narasimhan2007geometric}. 
    \subsection{Applications of WSPDs}
    
%    Well-separated pair decompositions are used to solve a wide variety of proximity problems. In this work, we consider the following five applications of \textsf{WSPD}s.
    
    \begin{itemize}
      
    \item \textsc{Construction of $t$-Spanners.} Given a pointset $P$ and $t\geq 1$, a $t$-spanner on $P$ is a Euclidean geometric graph $G$ on $P$ such that for every pair of points $p,q \in P$, the length of the shortest-path between $p,q$ in $G$ is at most $t$ times the Euclidean distance between them. %Such spanners are commonly used in robotics and computer networks for efficient geometric network design.
%    Interestingly, \textsf{WSPD}s can be used to construct $t$-spanners in $d$-space. 
%Our implementation is restricted to $2$-space. 
Refer to Algorithm~\ref{alg:spanner}. It returns the set of spanner edges and can be implemented to run in $O(n\log n)$ time~\cite{narasimhan2007geometric}.
    
	\begin{algorithm}[H]
	\caption{: \textsc{Construct-$t$-Spanner}($P,t>1$)}

	%	\begin{algorithmic}[1]
	%\STATE 
	
	     Let $s = 4(t+1)/(t-1)$. Construct a \textsf{WSPD} of $P$ with separation ratio $s$.
	    For every pair $(A_i,B_i)$ of the decomposition do the following: include the edge $\{a_i,b_i\}$ in the spanner where $a_i$ is an arbitrary point in $A_i$ and $b_i$ is an arbitrary point in $B_i$.
	 	\label{alg:spanner}
	\end{algorithm}

    \item \textsc{Finding Closest Pairs.} The problem asks to find two distinct points of $P$ whose distance is minimum among the $C(n,2)$ point pairs. %It is a fundamental problem in computational geometry and many algorithms are  known for this particular problem.
    The idea of well-separatedness can be used to design an algorithm for this problem. See Algorithm~\ref{alg:closestpair}. %The algorithm first constructs a $2$-spanner using the previous algorithm and then uses the spanner edges to find the closest-pair. %The algorithm returns a closest-pair.
    It can be implemented to run in $O(n\log n)$ time~\cite{narasimhan2007geometric}.
    
    	\begin{algorithm}
    	\caption{: \textsc{ClosestPair}($P$)}

 	Construct a $2$-spanner using Algorithm~\ref{alg:spanner}.
	     Since the closest pair is connected by an edge of the spanner, find the pair by iterating over all the edges.
	%	\end{algorithmic}
	\label{alg:closestpair}
	\end{algorithm}
    
    \item \textsc{Finding $k$-Closest Pairs.} It is a generalization of the closest-pair problem. Given  a positive integer $k$ such that $k \leq C(n,2)$, the goal is to find the $k$ closest pairs among the $C(n,2)$ pairs. See Algorithm~\ref{alg:kclosestpair}. It can be implemented to run in $O(n\log n + k)$ time~\cite{narasimhan2007geometric}.%As evident, this algorithm is more involved than the one for finding closest-pair. 
    
    	\begin{algorithm}\caption{: \textsc{$k$-ClosestPairs}($P$)}

\begin{enumerate}

\item Create a \textsf{WSPD} with some $s>0$. For every pair $(A_i,B_i)$ in the decomposition, let $R(A_i)$ and $R(B_i)$ be the bounding boxes of $A_i$ and $B_i$, respectively. Further, by $|R(A_i)R(B_i)|$, we denote the minimum distance between the two bounding-boxes $R(A_i), R(B_i)$. Renumber the $m$ pairs in the decomposition such that $|R(A_1)R(B_1)| \leq |R(A_2)R(B_2)| \leq \ldots \leq |R(A_m)R(B_m)|.$
    	    
    	    \item Compute the smallest integer $\ell \geq 1$, such that $\sum_{i=1}^\ell |A_i|\cdot|B_i| \geq k$.
    	    
    	    \item Let $r:= |R(A_\ell)R(B_\ell)|$.
    	    
    	    \item Compute the integer $\ell'$, which is defined as the number of indices with $1 \leq i \leq m$, such that $|R(A_i)R(B_i)| \leq (1+4/s)r$.
    	    
    	    \item Compute the set $L$ consisting of all pairs $\{p,q\}$ for which there is an index $i$ with $1 \leq i \leq \ell'$, such that $p \in A_i, q \in B_i$ or $q \in A_i, p \in B_i$.
    	    
    	    \item Compute and return the $k$ smallest distances determined by the pairs in the set $L$.
\end{enumerate}    	    
	\label{alg:kclosestpair}
	\end{algorithm}

    \item \textsc{Finding All-Nearest Neighbors.} In this problem, for every point $p$ in $P$, we need to find its nearest neighbor $q$ in $P\setminus \{p\}$. Refer to Algorithm~\ref{alg:ann} for a description of the algorithm. It can be implemented to run in $O(n\log n)$ time~\cite{narasimhan2007geometric}. 
    % to see how \textsf{WSPD}s can be used to design an algorithm for this problem.
    
\begin{algorithm}\caption{: \textsc{AllNearestNeighbors}($P$)}

    Choose $s>2$ and obtain the pairs of \textsf{WSPD}. For every $p \in P$, compute its nearest neighbor  in the following way:  
%\begin{enumerate}[label=(\alph*)]
     Find all such pairs of the \textsf{WSPD}, for which at least one of their sets is a singleton containing $p$.
     For every such pair $(A_i,B_i)$, if $A_i = \{p\}$, then $S_p = S_p \cup B_i$, else if $B_i = \{p\}$, then $S_p = S_p \cup A_i$. The nearest neighbor of $p$ is the  point in $S_p$ closest to $p$ (found by exhaustive search). 
%\end{enumerate}

	\label{alg:ann}
	\end{algorithm}
    
    \item \textsc{$t$-Approximate Minimum Spanning Trees}. Let $t>1$, be a real number. A tree connecting the points of $P$ is called a $t$-approximate minimum spanning tree of $P$, if its weight is at most $t$ times the weight of the Euclidean minimum spanning tree of $P$. Refer to Algorithm~\ref{alg:appxmst}. In $d$-space, it runs in $O(n\log n + n/(t-1)^d )$ time~\cite{narasimhan2007geometric}.
    
    \begin{algorithm}[H]\caption{: \textsc{$t$-ApproximateMinimumSpanningTree}($P, t > 1$)}

     Compute the $t$-spanner $G$ using Algorithm~\ref{alg:spanner}. Using Prim's algorithm compute a minimum spanning tree $T$ of $G$. Return $T$.

	\label{alg:appxmst}
	\end{algorithm}
\end{itemize}

%\section{Implementation}

%We have implemented our work using the \textsf{JSXGraph} library. Some code segments have been borrowed from the tool presented in \cite{anderson2021interactive}. %This tool will be permanently maintained by the third author  at \url{https://wisno33.github.io/VisualizingWSPDsAndTheirApplications/}.
%WSPDs are also covered in many computational geometry courses across the globe. In this connection, we believe that our tool will come in handy for teaching. 

%\medskip
%\noindent

%\textbf{Declaration.} Our tool is hosted on \texttt{GitHub}. If our paper gets accepted, we can demonstrate the tool live online or asynchronously using a video.

%\newpage

% \begin{sidewaysfigure}[ht]
%     \centering
%     \includegraphics[scale=0.3]{1} \\
%     \caption{A screenshot of page 1 of the tool where points are added.}
%     \label{fig:f1}
% \end{sidewaysfigure}

% \begin{sidewaysfigure}[ht]
%     \centering
%   \includegraphics[scale=0.3]{2}
%     \caption{A screenshot of page 2 of the tool where an algorithm is chosen for spanner construction.}
%     \label{fig:f2}
% \end{sidewaysfigure}
 
 %\clearpage 
 
%\bibliographystyle{plain}
\bibliography{a.bib}

\end{document}